\documentclass[12pt,twoside]{article}
\usepackage{fleqn,espcrc1,pstricks}

\usepackage{graphicx}
\usepackage[figuresright]{rotating}

\newcommand{\AmS}{{\protect\the\textfont2
  A\kern-.1667em\lower.5ex\hbox{M}\kern-.125emS}}

\title{
Dileptons and photons from central heavy-ion collisions at CERN-SPS
}

\author{
B. K\"ampfer\address[FZR]{Forschungszentrum Rossendorf,
PF 510119, D-01314 Dresden, Germany}\thanks{Supported by 
BMBF grant 06DR921.},
K. Gallmeister\addressmark[FZR]\thanks{Presently at University Giessen,
Germany.},
O.P. Pavlenko\addressmark[FZR]\address[ITP]{
Institute for Theoretical Physics, 
252143 Kiev - 143, Ukraine}
and
C. Gale\address{
Physics Department, McGill University, Montreal, QC, H3A 2T8, Canada}
}
      
\begin{document}

\maketitle

\begin{abstract}
A unique parameterization of secondary (thermal) dilepton and photon
yields in heavy-ion experiments at CERN-SPS is proposed.
%This parameterization resembles the lowest-order Born rates
%in a quark-gluon plasma.
Adding those thermal yields to  background contributions 
%(hadronic cocktail, Drell-Yan, correlated semileptonic decays of open charm) 
the spectral shapes of the
CERES/NA45, NA38, NA50, HELIOS/3 and WA98 data 
from experiments with lead and sulfur beams can be well
described.
\end{abstract}

\section{INTRODUCTION}

Electromagnetic radiation 
represents penetrating probes mapping out the full space-time
history of strongly interacting matter
in the course of heavy-ion collisions.
Particularly interesting is the stage of maximum temperature, where a
quark-gluon plasma might be formed.
%There are preliminary indications that already at CERN-SPS 
%a quark-gluon plasma is created.
% are summarized in \cite{Heinz_Jacob}.
Various properties of the quark-gluon plasma are calculable from
first principles exploiting numerical lattice QCD techniques. 
The results, such as thermodynamical properties, 
can be interpreted within
quasi-particle models \cite{quasi-particle}.
Other quantities, such as the electromagnetic emission
rates, are accessible within perturbation theory
\cite{Aurenche}.
%whichs validity near the confinement temperature $T_c \approx 170$ MeV
%is questionable. 

The motivation to measure electromagnetic signals
in ultralrelativistic heavy-ion collisions 
is based on the hope to have a direct access
to the hottest and densest stages of matter (cf.\ \cite{KOG}).
According to the present understanding of the dynamics of central heavy-ion
collisions at CERN-SPS, the temperature range covered
is $T_c \pm 50$ MeV 
(where $T_c \approx 170$ MeV is the confinement temperature).
In this regime the electromagnetic emission rates from 
deconfined and confined hadronic matter are fairly similar
(cf.\ \cite{KLS,Li_Gale}). Indeed, taking into account the 
empirical electromagnetic formfactors from the reaction 
$e^+ e^- \to hadrons$ and including vector - axial vector mixing
\cite{Huang} one gets a dilepton rate which coincides with the
lowest-order $q \bar q$ rate for invariant masses $M \ge 1.1$ GeV
\cite{Li_Gale}.
Below 1.1 GeV the dilepton spectrum of hadronic matter is determined
by the sharp $\phi$ and $\omega$ peaks and the broad $\rho$ peak.
Hadronic reactions broaden the $\rho$ peak so much that
the net dilepton spectrum again resembles that from the 
$q \bar q$ rate
down to $M \approx 300$ MeV \cite{Rapp_Wambach}.
It is therefore tempting to entirely describe the spectra at $M > 300$ MeV
by the lowest-order $q \bar q$ rate, thus treating it as a convenient
parameterization. 
%Below 300 MeV anyhow
%the Dalitz decays of hadrons dominate the spectra.
% so that the
%pecularities of the quark-gluon plasma rate \cite{Peshier_Thoma}
%probably are not relevant for interpreting the spectra.
Analog arguments are not available for the real photon rate.
Therefore, we use as a working hypothesis the lowest-order
Born rates of the processes $q \bar q \to \gamma g$ and $q g \to
\gamma g$ for thermal photons.

Many hadron observables at CERN-SPS energies can be described 
by thermal models.
%(cf.\ \cite{Redlich} for a survey and \cite{R.Rapp,C.Greiner}).
In line with this finding we will describe the secondary real and
virtual photons by a thermal source which adds to
the Drell-Yan-like yields and hadron decay
contributions (hadronic cocktail and correlated
semi-leptonic decays of open charm mesons).
%While the hadronic cocktails for the low-mass dilepton experiments
%are delivered by the CERES collaboration, the determination of
%the Drell-Yan like hard yields and the open charm contribution
%need separate efforts; the hadronic decay contributions to the
%photon spectra by the WA80 and WA98 collaborations are already 
%subtracted.

\section{ANALYSES OF EXPERIMENTAL DATA}

By now the following experimental data are available:
(i) lead beam at 158 A$\cdot$GeV and Au, Pb targets:
CERES (low-mass $e^+ e^-$),
NA50 (intermediate-mass $\mu^+ \mu^-$),
WA98 (direct photons),
(ii) lead beam at 40 A$\cdot$GeV and Au target:
CERES (low-mass $e^+ e^-$),
(iii) sulfur beam at 200 A$\cdot$GeV and Au, U, W targets: 
CERES (low-mass $e^+ e^-$),
NA38 (intermediate-mass $\mu^+ \mu^-$),
HELIOS/3 (low- and intermediate-mass $\mu^+ \mu^-$),
WA80 (upper bounds on direct photons).
In addition, data with proton beams at these energies and
also 450 GeV for various target nuclei are at our disposal.

\subsection{A simple parameterization of the thermal source}

The lowest-order rates are in Boltzmann approximation
\begin{equation}
\frac{dN_{\gamma*}}{d^4 x d^4Q} 
% & = & \frac{5 \alpha^2}{36 \pi^4}
\propto
\exp 
\left\{ - \frac{Q \cdot u}{T} \right\},
% \label{d_rates_1} \\
\quad
E \frac{dN_\gamma}{d^4 x \, d^3 q} 
% & = & \frac 59 \frac{\alpha \alpha_s}{2 \pi^2} 
\propto
T^2 
\exp \left\{ - \frac{q \cdot u}{T} \right\}
\log \left[ 1 + \frac{\kappa (q \cdot u)}{\alpha_s T} \right],
\label{rates_1}
\end{equation} 
where $Q$ ($q$) is the virtual (real) photon's 4-momentum;
effects of a finite chemical potential $\mu$ can be absorbed in an
additional normalization factor. 
Note the necessary occurrence of the medium's 4-velocity $u$ to build
up Lorentz-invariant rates.

To reduce the number of parameters in the space-time
integration, we replace in (\ref{rates_1})  
$T(t, \vec x) \to \langle T \rangle \equiv T_{\rm eff}$, 
$u(t, \vec x) \to \langle u \rangle$ and
$\int dt \, d^3x \to \int dt V(t) \to N_{\rm eff}$ yielding
\begin{equation}
\frac{dN_{\gamma *}}{d^4 Q} 
% & = &
\propto 
N_{\rm eff} 
% \frac{5 \alpha^2}{36 \pi^4} 
\exp \left\{ 
- \frac{M_\perp \cosh (Y  - Y_{\rm cms})}{T_{\rm eff}}\right\},
% \label{d_rates_2} \\
\quad
E \frac{dN_\gamma}{d^3 q}  
% & = &
\propto
N_{\rm eff} 
%\frac{3}{4 \pi} 
\int \cdots \exp \left\{ - \frac{q_\perp \cosh y }{T_{\rm eff}}
\cdots \right\}
\label{rates_2}
\end{equation} 
thus not relying on a particular model for 
$T(t, \vec x)$, $u(t, \vec x)$ and $\mu(t, \vec x)$
(for the explicit form of (\ref{rates_2}) 
cf.\ \cite{our_g_model,our_d_model}). 
This approach is in the spirit of the usual parameterization of the
transverse momentum spectra of hadrons by an exponential with
a slope parameter and a normalization factor.
As an example we show in Figure 1 our results for the invariant
mass spectra of dileptons and the momentum spectrum of photons
for the experiments of group (i)
(cf.\ \cite{our_g_model,our_d_model} for more details). 
Also the transverse momentum spectra of dileptons are described 
very well \cite{our_d_model}. The NA50 acceptance
is incorporated only in an approximate way here; the use of the correct
filter reproduces the data perfectly \cite{Capelli}.

A comparison of our model with the preliminary CERES data
\cite{CERES_40GeV}
of the experiment group (ii) is shown in Figure 2
for $T_{\rm eff} = 145$ MeV and a fairly large value of 
$N_{\rm eff}$.\footnote{
Smaller values of $T_{\rm eff}$ deliver a better description of the
spectral shape but need even larger $N_{\rm eff}$.}

Since the experiments of group (iii) cover a large rapidity range
an additional Gaussian smearing of the source in 
(\ref{rates_2}) needs to be included \cite{our_d_model}.
In Figure 3 the strength distributions of the thermal
dilepton source without and with smearing are displayed. 
As shown in \cite{our_d_model} 
the shape of the spectra can be well
described by  $T_{\rm eff} = 160 \cdots 170$ MeV, however,
different values of $N_{\rm eff}$ are needed. This might be attributed
to an inadequate rapidity distribution of our thermal source
and different centrality selections in the experiments.

\begin{figure}[h] %%%%%%%%%%%%%%%
\centering
\includegraphics[angle=-90,width=5cm]{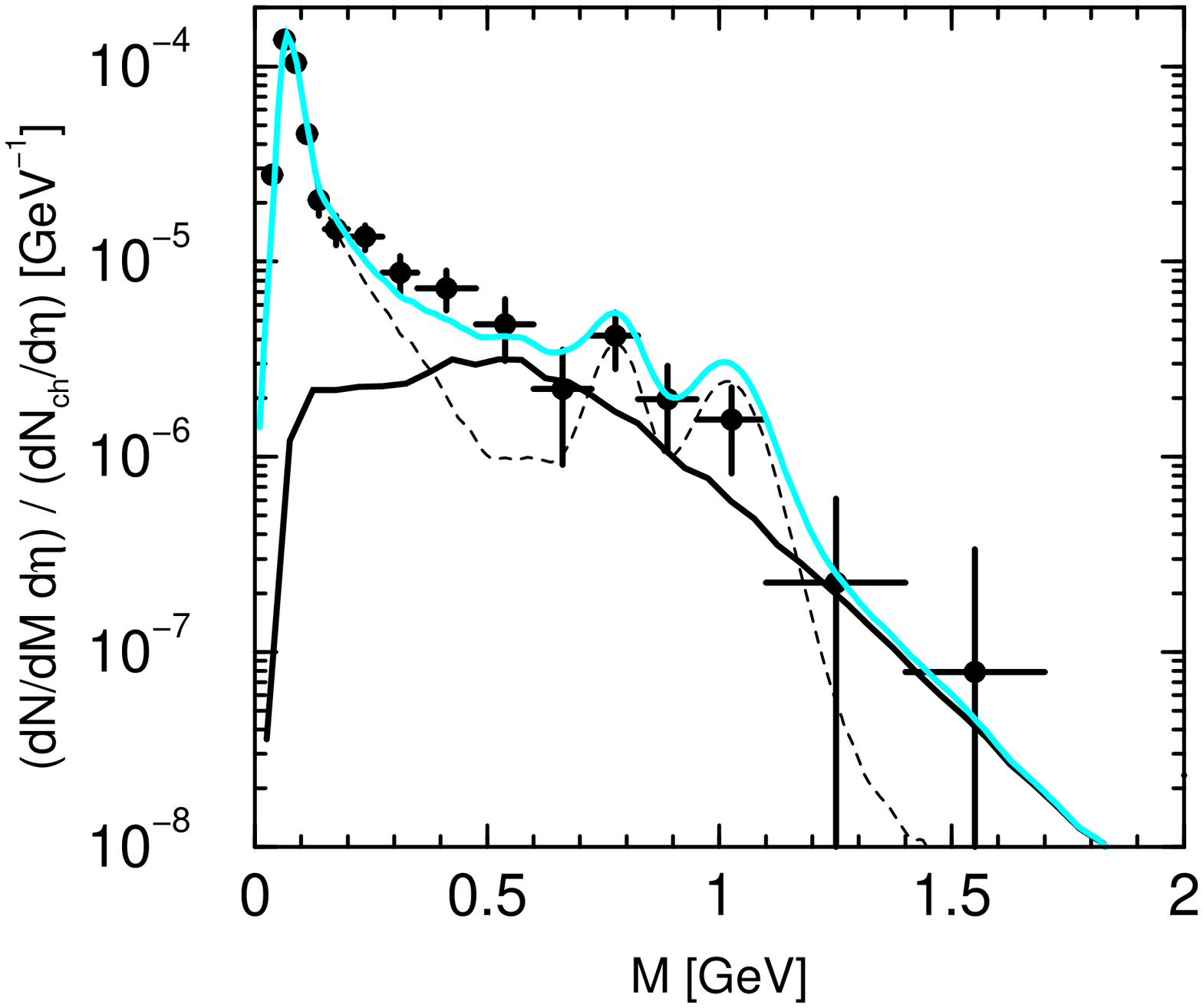}
\psline[linecolor=red]{->}(-3.5,-2.45)(-3.55,-1.55)
\rput(-2.9,-2.6){{\footnotesize thermal}}
\psline[linecolor=black]{->}(-3.0,-2.2)(-3.0,-1.9)
\rput(-2.7,-2.3){{\footnotesize cocktail}}
\hfill
\includegraphics[angle=-90,width=5cm]{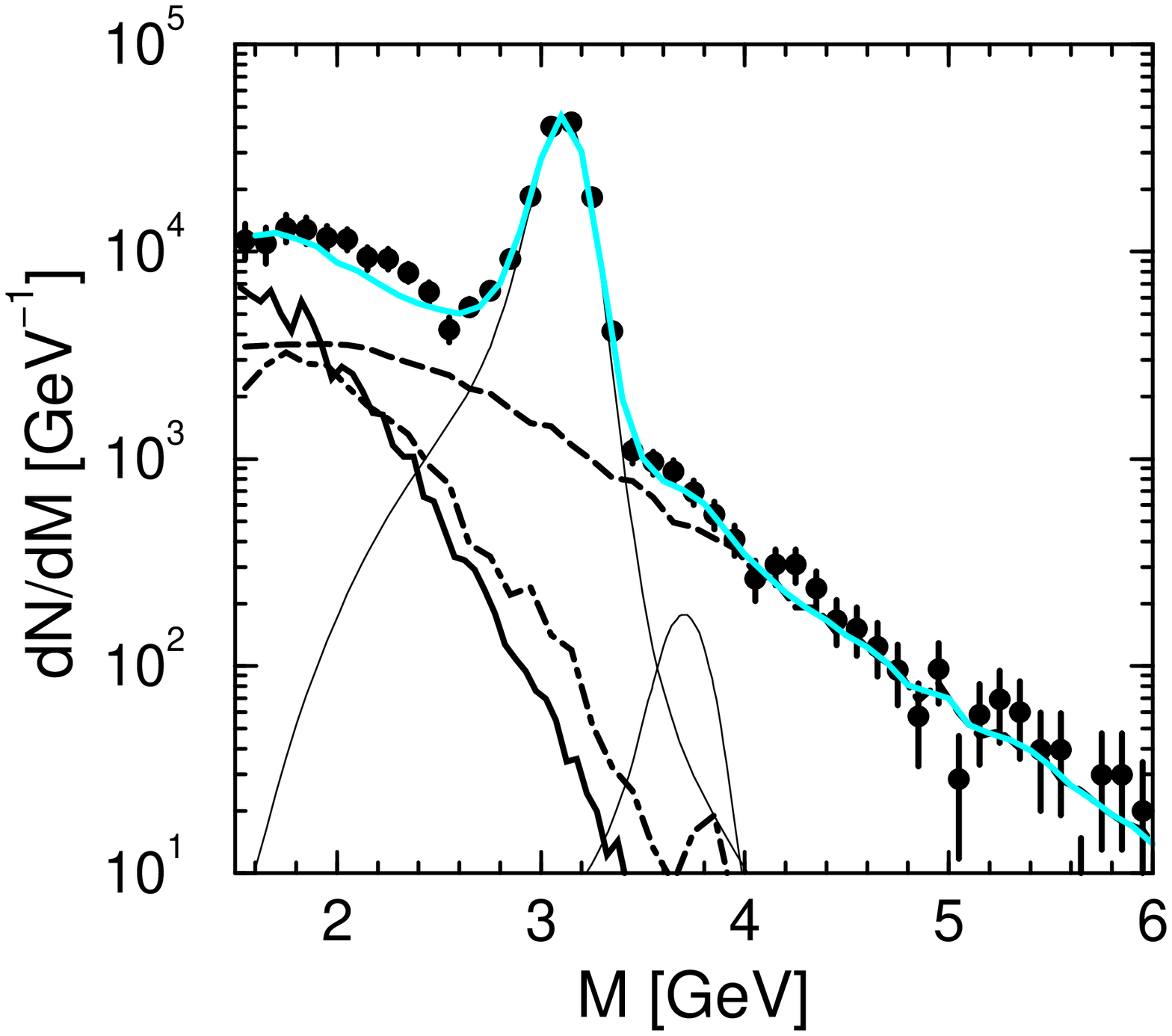}
\psline[linecolor=black]{->}(-2.3,-1.4)(-2.88,-1.68)
\rput(-1.5,-1.4){{\footnotesize Drell-Yan}}
\psline[linecolor=black]{->}(-3.1,-2.9)(-2.76,-2.66)
\rput(-3.6,-2.9){{\footnotesize charm}}
\psline[linecolor=red]{->}(-2.85,-3.41)(-2.65,-3.35)
\rput(-3.5,-3.4){{\footnotesize thermal}}
\hfill
\includegraphics[angle=-90,width=5cm]{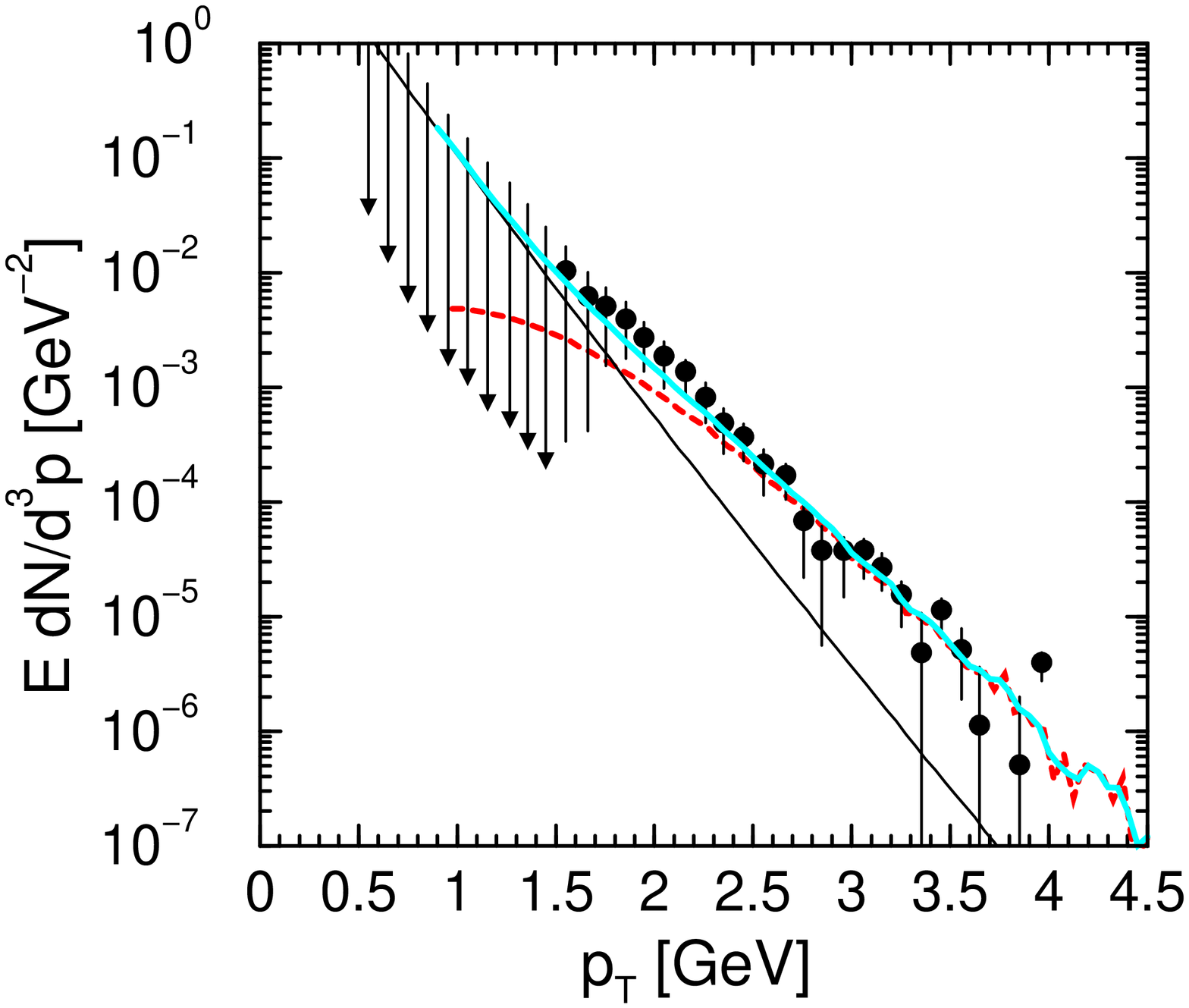}
\psline[linecolor=red]{->}(-2.6,-2.77)(-2.,-2.2)
\rput(-2.5,-2.9){{\footnotesize thermal}}
\rput(-2.5,-3.25){{\footnotesize ($v_r = 0.3$)}}
\psline[linecolor=black]{->}(-3.3,-2.3)(-2.95,-1.28)
\rput(-3.3,-2.5){{\footnotesize pQCD}}
~\\[-6mm]
\caption{
Comparison of our model with dilepton data 
(left panel for the CERES data \protect\cite{CERES};
middle panel for the NA50 data \protect\cite{NA50},
thin lines: parameterizations of the $J/\psi$ and $\psi'$)
and
the photon data (right panel for WA98 data \protect\cite{WA98};
hadron decay contributions are subtracted).
The thermal contribution
is characterized by the unique set of parameters 
$T_{\rm eff} = 170$ MeV and $N_{\rm eff} = 3.3 \times 10^4$ fm${}^4$.
Sum of all contributions: gray curves.}
\label{em_signals}
%\end{figure}
%\begin{figure}[h] %%%%%%%%%%%%%%%
\centering
\begin{minipage}[t]{5cm}
\includegraphics[angle=-90,width=5.2cm]{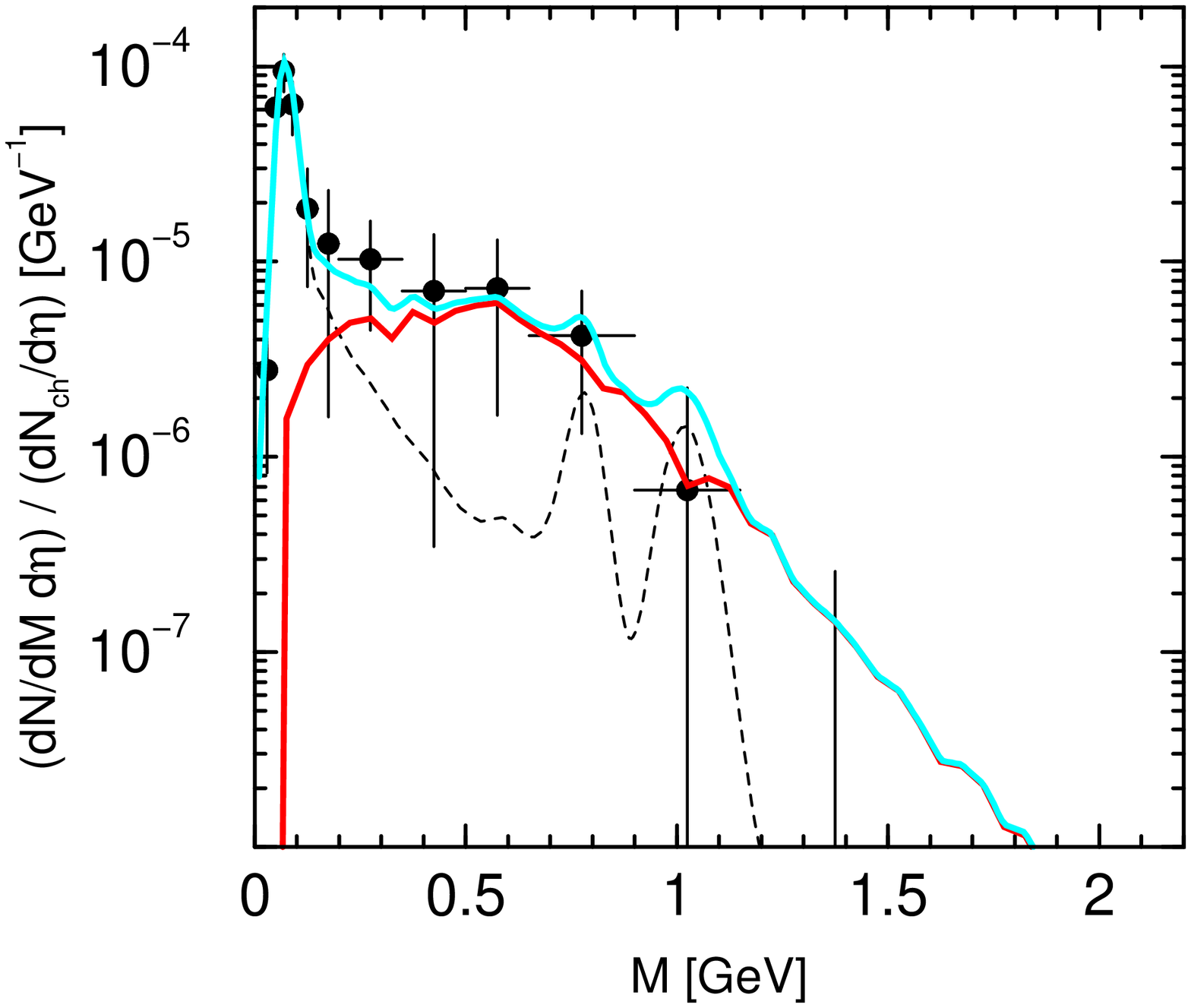}
\psline[linecolor=black]{->}(2.,1.7)(2.0,2.3)
\rput(2.,1.5){{\footnotesize cocktail}}
\psline[linecolor=red]{->}(2.8,3.56)(2.33,3.23)
\rput(3.5,3.6){{\footnotesize thermal}}
~\\[-13.1mm]
\caption{
Comparison of our model with preliminary CERES data 
\protect\cite{CERES_40GeV} of the lead beam run at 40 A$\cdot$GeV.}  
\label{40GeV_CERES}
\end{minipage}
\hfill
\begin{minipage}[t]{10.1cm}
\includegraphics[angle=-90,width=5cm]{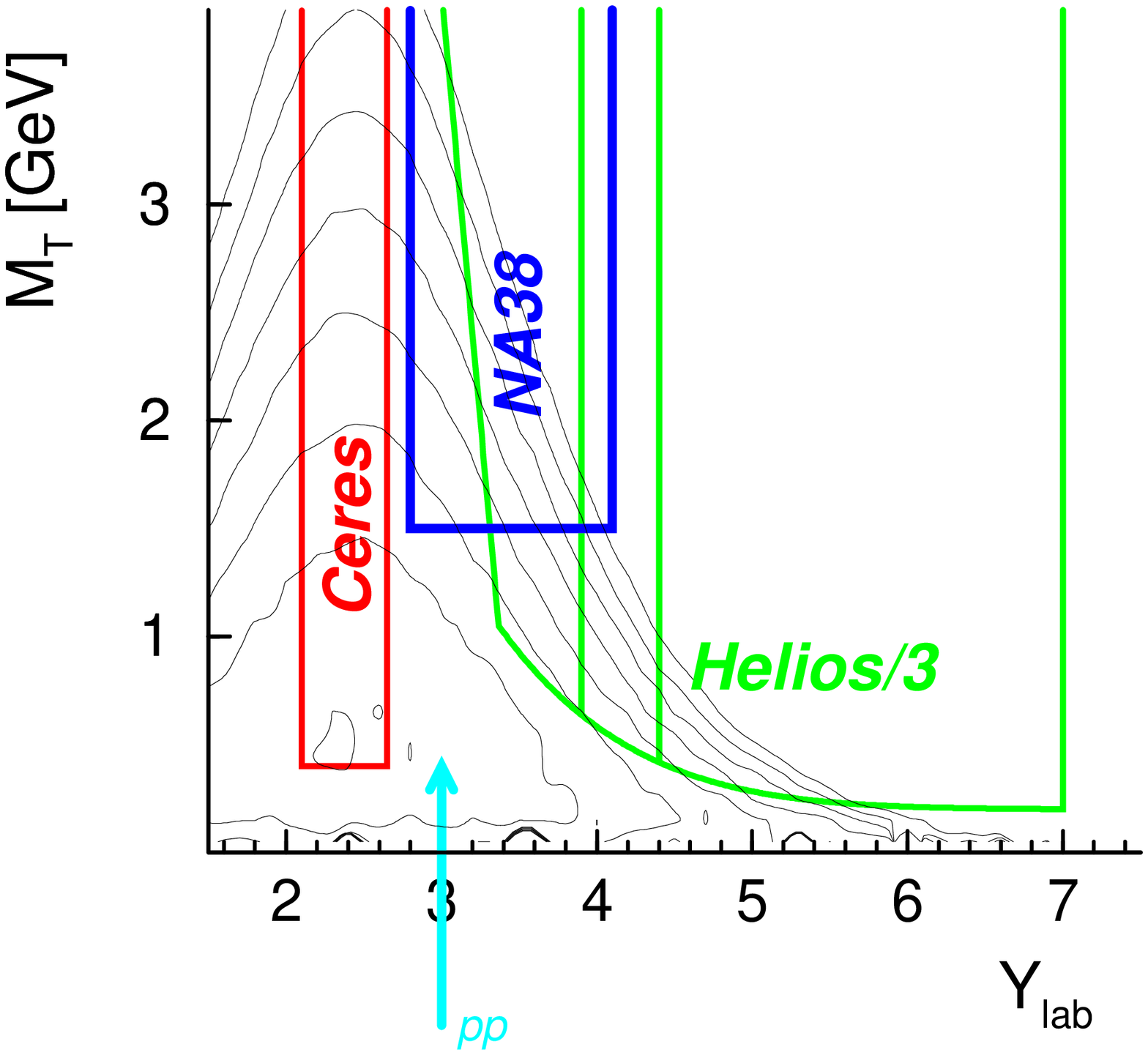}
\hfill
\includegraphics[angle=-90,width=5cm]{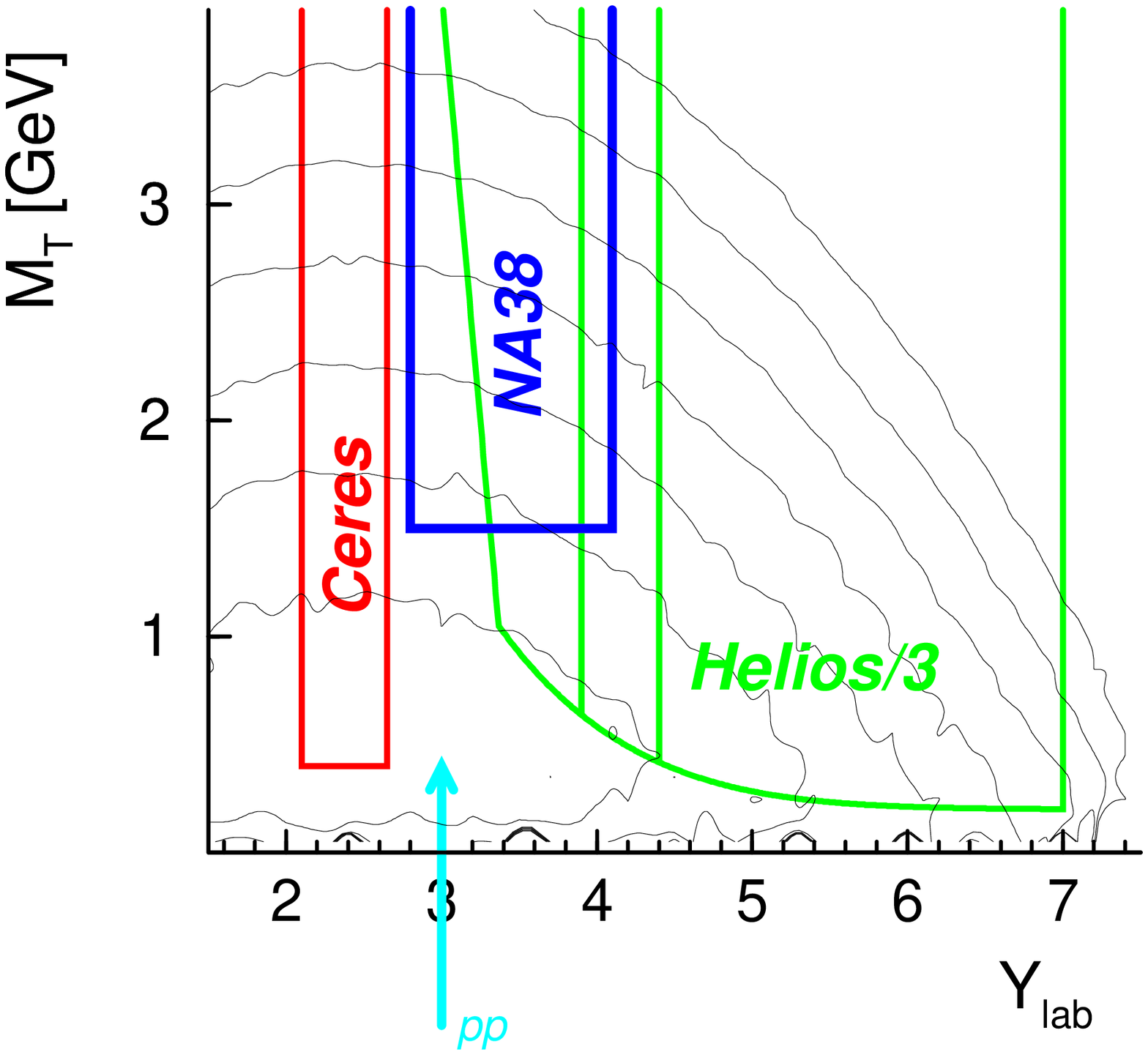}
~\\[-13.9mm]
\caption{
Contour plots of the thermal dilepton distribution in
rapidity vs.\ transverse mass space without (left panel) and
with (right panel) smearing in rapidity space for $T = 160$ MeV.
%The acceptance regions of the various experiments are displayed too.
}
\label{contours}
\end{minipage}
\end{figure}

\subsection{Time evolution effects}

By using a model for $T(t)$ and $V(t)$ and neglecting
spatial gradients one can equally well describe the dilepton 
spectra of the CERES and NA50 collaborations. As shown in
\cite{our_dileptons},
there is virtually no difference with 
the results displayed in Figure 1.
The needed initial temperature is 210 MeV, for a given freeze-out
temperature of 120 MeV.
 
\subsection{Background contributions}

While the analysis of the low-mass dileptons
uses a hadronic cocktail and is normalized
to the mean multiplicity of charged hadrons, the analyses of the 
intermediate-mass dileptons and photons need careful estimates
of the hard Drell-Yan-like yields and the open charm contribution.
We used the PYTHIA event generator with
charm quark mass of 1.5 GeV, intrinsic parton
transverse momentum distribution width 
%$\sqrt{\langle k_\perp^2 \rangle} = 
of 0.8 GeV and
photon cut-off parameter of 1 GeV. The 
parameter adjustments are done by comparing with $pp$ and $pA$
data, and then scaling to heavy-ion collisions. The possible subtleties
of this procedure are described in \cite{our_d_model,Gallmeister}.  

\section{CONCLUSIONS}

In summary we have demonstrated that the superposition
of background contributions and a thermal source of real and virtual
photons describes the shape of all data from central heavy-ion
collisions at CERN-SPS. We find
evidence for a large and long-lived thermal source
in lead beam experiments at 158 A$\cdot$GeV.
The normalization factor $N_{\rm eff}$ translates into a lifetime
of 23 fm/c, in accordance with \cite{Shuryak}, when assuming a
spatial volume of the fireball 1440 fm${}^3$ as found for a
radius of 7 fm. The space-time averaged temperature of
$T_{\rm eff} = 170$ MeV coincides with both the confinement
temperature and the temperature needed to describe the
hadron multiplicity ratios in thermal models.
Using a model for the temperature and volume
evolution of the fireball we find a maximum temperature of
210 MeV indicating an initial state 
%of strongly interacting matter
in the deconfinement region.
The sulfur beam data point to a similar value of $T_{\rm eff} = 160$ MeV, 
%but to different values of the thermal strength
%factor $N_{\rm eff}$ which might be attributed in part to different
%centrality selections in the experiments and in part to a 
%improper source strength distribution in rapidity space.
while the recent preliminary lead beam data at 40  A$\cdot$GeV
can be described by   $T_{\rm eff} \le 145$ MeV.

The present analysis strategy needs to be extended to 
understand the centrality dependence of the intermediate-mass
dileptons as measured by the NA38/50 collaborations \cite{Capelli}.
The results of the high-statistic data sampling of dileptons in the lead beam 
induced reactions at 158 A$\cdot$GeV by the CERES collaboration
is eagerly waited for to decide whether a smooth $q \bar q$-like rate
is still sufficient. We used it here as a convenient parameterization, and
this is not a substitute to a more detailed microscopic calculation.
% or a modified spectrum (e.g., with shifted
%resonance contributions) are needed.

The consistent description of the real and virtual photon data
makes an interpretation of the NA50 data by an anomalous open charm
enhancement unlikely. Nevertheless, the explicit measurement of the charm
yield, as envisaged by the NA60 collaboration, is needed
to arrive at a better understanding of the various dilepton sources. 

%%%%%%%%%%%%%%%%%%%%%%%%%%%%%%%%%%%%%%%%%%%%%%%%%%%%%%%%%%%%%%%%%%%%%%%%%

\end{document}